# A Knowledge-driven Business Process Analysis Canvas

(Draft ver. v01, 30 November 2021)

Michele Missikoff

IASI-CNR, Via dei Taurini 19, Rome, Italy
michele.missikoff@iasi.cnr.it

**Abstract**. Business process (BP) analysis represents a first key phase of information system development. It consists in the gathering of domain knowledge and its organization to be later used in the software development, and beyond (e.g., for Business Process Reengineering). The quality of the developed information system largely depends on how the BP analysis has been carried out and the quality of the produced requirement specification documents. Despite the fact that the issue is on the table for decades, business process analysis is still a critical phase of information systems development. One promising strategy is an early and more important involvement of business experts in the BP analysis. This paper presents a methodology that aims at an early involvement of business experts while providing a formal grounding that guarantees the quality of the produced specifications. To this end, we propose the Business Process Analysis Canvas, a knowledge framework organized in eight knowledge sections aimed at supporting the business expert in carrying out the analysis, eventually yielding a BP analysis Ontology.

## 1  Introduction

Business process (BP) analysis is a key activity for many strategic projects of an enterprise that go from the business process reengineering to information system development. When analyzing a BP, it cannot be considered in isolation with respect to other elements of the enterprise. Even if our initial focus is on a specific business process, we need to consider other related elements, such as documents, enterprise organization, roles and skills of the involved people. Conversely, if our focus is, e.g., on product innovation, then we are forced to reconsider the involved processes as well.

BP Analysis [1] is positioned in the preliminary phase of a software project. Software projects are among the most difficult engineering undertakings, difficult to be managed and carried out accomplishing a final product with the required characteristics, within the time and budget originally planned.

Despite the significant advances of Software Engineering, and specifically Requirement Engineering, software projects still face a number of problems, such as the ones listed below:

- The high cost of software production
- Slow reaction time of IT departments to the requests of new software solutions
- A serious shortage of software analysts and programmers
- The high number of software projects exhibiting poor achievement or even total failure
- Released software applications that are never free of bugs, with some forms of unreliable (and unpredictable) behaviour
- And, last but not least, the big problem of business/IT alignment [2], i.e., the misalignment between business needs and the services offered by the information system, that is still on the table.

One of the main causes of such problems resides in the first phase of the software development project, i.e., in the collection of business and user requirements that often yield poor requirements specifications, modeling, and validation. Among the main causes are the insufficient involvement of business people and the poor quality of the requirement specification documents.

In the paper, we address BP analysis with a knowledge-driven approach, starting with the gathering of domain knowledge, typically through interviews, existing documents and observation to produce a number of knowledge artifacts. Then, the knowledge is represented as a collection of models to be shared among domain experts and stakeholders for collaboration and validation.

The proposed methodology is characterised by the following points: (i) ease of use for the business experts who are moved at the center of the stage; (ii) well defined guidelines, with a progression from informal to formal (inspired by some early ideas [3]); (iii) knowledge-driven approach based on a solid formal grounding. The points may sound contradictory, since rigid formality reduces the acceptance of business people. Our proposal tries to solve this contradiction.

The collected knowledge is organized according to a grid, inspired by the Business Model Canvas [4], named Business Process Analysis Canvas (BPA Canvas). With some key differences: primarily the scope, in fact the former aims at organising the whole enterprise space, while our proposal is more focused, addressing in a more detailed fashion a specific application domain with the objective of business process analysis. Furthermore, the proposed method requires the production of a number of knowledge artefacts, providing clear definitions of their structure and guidelines for their production. Finally, while the Business Model Canvas is inherently informal, the BPA Canvas has a formal grounding and aims, at the end of the analysis, to release a BP ontology as a final formal artefact.

We are aware that, traditionally, BP Analysis (BPA) is a territory of business experts, who adopt methodologies that are mainly descriptive, without a rigorous approach for the activities and the released documents [5]. The informal nature of the produced documents, often containing imprecise statements or missing information, are among the causes of the encountered pitfalls, in particular in the development of enterprise information systems. Several solutions, in particular in the context of Requirement Engineering, have been proposed, but yet with a limited success [6]. According to Standish

Group's Annual Chaos report[1] (2020), 69% of software projects (based on the analysis of 50,000 projects globally) end in partial or total failure.

To solve the issue, one idea was to adopt knowledge-based methods, centered on business ontologies [7] but, as we will see in the Related Work Section, such methods appeared too complex and where not well received by the business community. Today there is not yet a winning solution for business experts to build and manage formal specifications of a business scenario.

The BPA Canvas proposes a methodology conceived to ease the adoption of a rigorous analysis method by business experts. It consists in a progressive construction of knowledge artefacts, in a sequence that starts from simple, narrative models, and then proceeds building semantically richer models, eventually achieving an ontology of the business scenario. All the models, except the final ontology, can be built without specific knowledge engineering competences.

The rest of the paper is organized as follows. Section 2 describes the BPA Canvas methodology, then Section 3 illustrates the methodology by means of a running example. The next Section 4 focuses on the last steps to achieve the BPA Ontology, followed by Section 5 that provides a short review of the literature in the area of knowledge management for business process analysis. Finally, Section 5 reports the conclusions and some lines for future research.

## 2 The Business Process Analysis Canvas

In this section, we introduce the main ideas of the Business Process Analysis Canvas, its knowledge artefacts and the related methodology aimed at guiding business experts in collecting and organising the knowledge of the application domain.

With respect to the business process modelling methods available in the literature, the BPA Canvas has not the objective of drawing process diagrams, conversely it intends to collect all the knowledge required to build a first static model. The actual diagramming is postponed to a subsequent phase, with the assumption that a rigorous and detailed structural knowledge-base about the BP will ease the design task and improve the quality of the produced process flow diagrams.

### 2.1 The BPA Canvas layout

The BPA Canvas is organised in eight knowledge sections that hold different kinds of knowledge artefacts, i.e., models of the given business domain. The models that represent the enterprise knowledge can assume various forms, with different levels of details and formality. In particular, we have: (i) *plain text*, a narrative form of knowledge representation; (ii) *structured text*, e.g., itemised lists (bullet points, numbered lists, etc.) that collect and organise short statements; (iii) *tables*, typically providing a systematic visualization of knowledge items; (iv) *diagrams*, where the knowledge is graphically represented, according to a given standard; (v) a formal representation of the business domain by means of a *BP Ontology*. Figure 1 shows the layout of the eight sections of the BPA Canvas.

---

[1] https://hennyportman.files.wordpress.com/2021/01/project-success-qrc-standish-group-chaos-report-2020.pdf

**Fig. 1**. BPA Canvas layout.

The knowledge sections of the BP canvas are:

a) **BP Signature**. The first knowledge artefact, in the form of a table, aimed at providing a synthetic description of the business process, gathering key information about it.
b) **BP Statement**. A preliminary plain text description of the business process, and its business scenario, described in general terms (i.e., at an intensional level).
c) **BP User story**. One or more plain text descriptions of exemplar executions of the BP (i.e., at an extensional level). In essence, it represents one or more instance of the BP Statement.
d) **Actor, Activity, Outcome (AAO) Matrix**. It represents the output of a first linguistic analysis of the BP Statement.
e) **PB Glossary**. A collection of terms, with their descriptions, that characterise the BP domain.
f) **OPAAL Lexicon**. This is a structured terminology that provides a first semantic tagging of the key terms used in the previous structures, according to the following categories: *Object, Process, Actor, Attribute, Link*.
g) **UML Class Diagram**. The construction of the UML Class Diagram (CD) starts form the knowledge collected so far, modeling it in a graphical form. Such a graphical representation is particularly useful for sharing the knowledge among people.
h) **BPA Ontology**. This is a formal representation of the analysed business process. It is the final knowledge artefact of the methodology.

## 2.2 The BPA Canvas methodology

The methodology, that will be applied to an example in the next section, suggests to start specifying the *BP Signature*, and then continue with *the BP Statement* and a number of *User Stories* (one for each business case). These models are built using plain text descriptions, easily provided by domain experts. Then, the *AAO Matrix* requires a first linguistic analysis of the collected knowledge, extracting simple triples: <*subject, verb, direct/indirect object*> from the given texts. In parallel, we start to build the *BP Glossary* that contains all the terms used in above knowledge artefacts, together with their descriptions. The Glossary represents a solid reference point, very useful when the picture gets large and complicated. Then we have the first semantic step: we need to classify the terminology according to the five categories of the *OPAAL (Object, Process, Actor, Attribute, Link) Lexicon*. The following step consists in creating *UML Class Diagrams*, starting from the content of the Lexicon and then the BPA Ontology that formalises the whole picture. We presented the BPA Canvas in a sequence, but in carrying out the analysis we would rather proceed in a spiral way, going back and forth to keep aligned and consistent the various models.

The above knowledge artefacts are sufficiently intuitive and can be built by business experts without specific technical competences (and, after a suitable training, without the help of knowledge specialists). The last step of the methodology consists in the construction of the *BPA Ontology* that models in formal terms (by using an ontology language, such as OWL) all the knowledge acquired in the previous steps. The BPA ontology requires the intervention of an ontology engineer.

## 3 Applying the BPA Canvas: a running example

As anticipated, the BPA Canvas focuses on the structural elements of the BP, where tasks, activities, operations are considered as entities to be linked with the other business elements (document, actors, etc.). Therefore, according the philosophy of an incremental knowledge modeling, we postpone the intricacy of the business logic and the formal modeling of the temporal sequencing tasks to a later moment (that will be addressed in a next paper).

### 3.1 A running example

The example chosen to illustrate the BPA Canvas consists in a home delivery pizza business. The aim is to show the progression in building the knowledge artefacts in a stepwise fashion to tame the complexity and formality of the methodology, until the BPA ontology is eventually produced.

**BP Signature.** The Table 1 represents the first knowledge structure of the pizza shop BP. It is then followed by the BP Statement and the User Stories (just one in our case).

Table 1. Pizza shop BP Signature.

| Knowledge items | Content |
|---|---|
| **BP Name** | Home Pizza Delivery |
| **Trigger** | Order Arrived |
| **Key Actors** | Customer, Cook, Delivery Boy |
| **Key objects** | Order, Dough, Pizza, Delivery Vehicle |
| **Input** | Order |
| **Objective** | Cook and deliver pizzas to customers |
| **Output** | Pizzas Delivered, Customer happy |

**BP Statement.** The text of the BP Statement is the synthesis of an interview to a (fictitious) pizza shop owner, whose business has name *PizzaPazza*.

*My business, PizzaPazza, is a home delivery pizza shop. The customer fills in the order and then submits it to the shop, with the payment, by using our Web site. Making good pizzas requires good quality dough, produced in-house, and a careful baking of the pizza. To make clients happy, we need to quickly fulfil the order and the delivery boy needs to know streets and how to speedily reach the customer's address.*

**BP User story.** This text reports a specific execution of the BP, i.e., it represents an instance of the PizzaShop BP. If necessary, more than one User story is reported, to represent various use cases (points of view.)

*Mary connects to the PizzaPazza Web site and places his order of two Napoli pizzas, providing also the payment. On the arrival of Mary's order at PizzaPazza, John, the cook, puts the order on the worklist. When the Mary's turn arrives, John prepares the ordered pizzas, cooks them, and then alerts the delivery boy Ed to come and pick up the pizzas. Thus, Ed collects the pizzas and starts his delivery trip, eventually achieving the delivery to Mary's home.*

The first three knowledge artefacts represent an important, but intuitive, starting point to approach the subsequent semantic analysis of the business scenario.

### 3.2 Semantic analysis of the BP Statement and User stories

The semantic analysis starts from the free-form text to extract a first structured knowledge artefact: *AAO Matrix*. As anticipated, it consists of triples representing independent clauses, organised in a matrix with three columns, with headers: Actor, Action, Outcome (AAO). The goal of the AAO Matrix is to seize from text the key knowledge about who (Actor) is doing what (Actions) yielding what results (Outcome).

In essence, according to linguistic theory, the text is analysed to extract triples formed by a subject noun phrase (S/NP), a verb phrase (V/VP) and a direct or indirect object noun phrase (DO-I/NP), following pattern: (S/NP, V/VP, DO-I/NP). Below the AAO Matrix in our example.

Table 2. AAO Matrix of Pizza shop.

| Actor | Action | Outcome |
|---|---|---|
| customer | filling and submitting | order |
| pizzaShop | receiving | order |
|  | making | pizza |
|  | producing | dough |
|  | baking | pizza |
| deliveryBoy | collecting | pizza |
|  | delivering | pizza |
| customer | appraising | service |

In the Action column, actions are represented using the gerund form instead of the more 'technical' stemming. Furthermore, clauses are represented in active verbal form, therefore if in the text we have a passive form (e.g., the order is issued by the customer), in building the triple we need to turn it into an active form (customer issuing an order).

### 3.3 Building the BP Glossary

This knowledge artefact is built starting from the textual models that have been produced so far. It is created extracting from the texts the relevant terminology, i.e., the terms that represents entities, attributes, and activities characterising the analysed business domain. For each term, a short description is provided and, if the case, one or more synonyms. Below an excerpt of the Pizza Shop Glossary (the descriptions have been derived from The Free Dictionary).

Table 3. PizzaShop Glossary.

| Term | Synonym | Description |
|---|---|---|
| … | | |
| **Customer** | Client | One who buys goods or services from a store or business. |
| **Cooking** | Baking | To cook food with dry heat, especially in an oven. |
| **DeliveryBoy** | Rider | One that performs the act of conveying or delivering. |
| **Order** | Purchase | A request made by a customer at a pizza shop for food |
| **PizzaKind** |  | Different types of pizza the customer can chose to order |
| … | | |

### 3.4 Building the BP Semantic Lexicon

In this step, we generate a lexicon that organises the terms into five semantic categories according to the OPAAL scheme.

(i) *Object*: any passive entity with a lifecycle that follows to the CRUDA paradigm, i.e., the traditional *Create, Read, Update, Delete* [8], to which we add *Archive* that is particularly relevant in business processes;

(ii) *Process:* any form of activity, function, operation aimed at enacting CRUDA operations on one or more business entity (Actor or Object);

(iii) *Actor:* any active entity involved in one or more processes;

(iv) *Attribute*: a property (simple or complex) associated to one or more of the former concepts;

(v) *Link:* this section captures domain relationships organising the terminology in pairs where the terms are semantically related [9]. The links are of two different sorts: structural links that represent, e.g., subClassOf or a partOf, and functional links, where the implied relationships is a domain action (e.g., cooking).

Table 4 reports an excerpt of the PizzaShop OPAAL Lexicon. Please note that here we do not mean to be complete, the reported structures have mainly an illustrative purpose.

**Table 4.** The OPAAL Lexicon of Pizza shop.

| Categories | Business terminology |
|---|---|
| **Object** | Order, Pizza, Margherita, Base, Topping, Address, … |
| **Process** | Cooking, MakingDough, PlacingOrder, AcceptingOrder, DeliveringPizza, ReceivingPizza, … |
| **Actor** | PizzaShop, Customer, Cook, DeliveryBoy, … |
| **Attribute** | Price, Quantity, PizzaKind, … |
| **Link** | *Structural*. Order-Pizza, Customer-Address, Pizza-Margherita,… |
| | *Functional*. Customer-Order, DeliveryBoy-Pizza, PizzaShop-Order, Customer-Pizza, PizzaShop-Pizza,… |

To better clarify the elements of the Table 4, we provide a formal account of its content, introducing five predicates, each of which corresponds to a row of the table.

- *object(x),* evaluate true if *x* is an object;
- *process(x)*, evaluate true if *x* is an activity, an operation, a task, a process;
- *actor(x)*, evaluate true if *x* is an actor;
- *attribute(x)*, evaluate true if *x* is an attribute;
- *linked(x,y)*, evaluate true if, in the given application domain, the concept represented by *x* is semantically related to the concepts *y*.

Assuming that we have the full application lexicon $\mathcal{T}$ that gathers all the terms used to describe the given business domain, then we define four subsets of $\mathcal{T}$:

$$O \subseteq \{o \in \mathcal{T} : object(o)\};$$
$$P \subseteq \{p \in \mathcal{T} : process(p)\};$$
$$A \subseteq \{a \in \mathcal{T} : actor(a)\};$$
$$AT \subseteq \{t \in \mathcal{T} : attribute(t)\}$$

and the relation *L*:

$$L = \{ (x,y) \in \mathcal{T} \times \mathcal{T} : linked(x,y) \wedge ( (actor(x) \wedge actor(y)) \vee (object(x) \wedge object(y)) \vee (actor(x) \wedge object(y)) ) \}$$

Please note that the above formalization does not intend to be complete, for instance constraints (e.g., disjointness) are not reported. Furthermore, for sake of brevity we left out the attributes that can be associated to all the entities. In the *Link* category we listed only the domain specific terms, giving for granted the general conceptual modeling constructs, such as *partOf, ISA (*the generalization operator*)*, etc.

## 4  Building Class Diagrams and the BPA Ontology

In this section we illustrate the guidelines for building the Class Diagrams and the final BP Ontology.

Starting from the above knowledge artefacts, and in particular from the OPAAL Lexicon, the next two artefacts consist in the UML-Class Diagrams (CD) [10] and the BPA Ontology of the Pizza shop. As anticipate, the eight sections of the BPA Canvas have been listed in a sequence, but their construction does not take place sequentially. In particular, in this section we carry out the building of the last two artefacts (diagrams and ontology) in parallel, presenting a selection of Class Diagrams together with the corresponding ontology fragments.

To build the CDs we start from the OPAAL Lexicon applying some rules sketchily reported below. To tackle the overall complexity, instead of building a single large diagram, we partition it in subdiagrams, adopting a partitioning criterion based on the different kinds of links. Diagram partitioning, based on well-known techniques rooted in Graph Theory, is not an easy job. In particular, it presents a number of problems when reconstructing a coherent global graph, especially if the semantics of edges and nodes is involved. We address the problems with ontology merging techniques [11]. Below the sketchy CD building rules:

- Class boxes are labelled with one of the terms in the **Object** or **Actor** categories.
- **Attribute** terms are listed within the box of the corresponding concept (not reported in the figures).
- Pairs of terms in the **Link** section are represented by arrows (with or without head) connecting two boxes. Such arrows include the structural relations:
  - *ISA*, if linking an object, an actor or a process with its more general concept.
  - *PartOf*, if linking an object, an actor or a process that is a component connected to a more complex assembly.

- *Action*, if the link is of a functional nature, connecting an actor with another actor or an object. The action name is one of those listed in the **Process** section (we recall that the term Process in OPAAL is more general than 'business process', including various behavioral notions, such as task, operation, action, activity, function).

Once a diagram has been built, we proceed in the construction of the corresponding BPA Ontology fragment, applying the following (sketchily reported) rules.
- *Object* and *Actor* boxes are modelled as OWL *classes*
- *Attributes* are modelled as *datatype Properties* (not reported in the example)
- *Arrows* are modelled as *Object Properties*, where *Domain* and *Range* are defined by the pair of boxes reported in the CD and the property name is the label of the link connecting the two boxes. Then:
  - If the domain (i.e., the source of the arrow) is an **Actor**, the *Object Property* represents an action on another Actor or Object (depending on the range).
  - If the domain is an **Object**, then the range is another object and the label is, for instance, *partOf*, or another relation among objects (e.g., *nextTo*).

For sake of space, we report only one structural and one functional CD, together with the corresponding ontology fragments (using a simplified Turtle syntax, e.g., omitting the namespaces). Below a *partOf* diagrams with the corresponding ontology fragment.

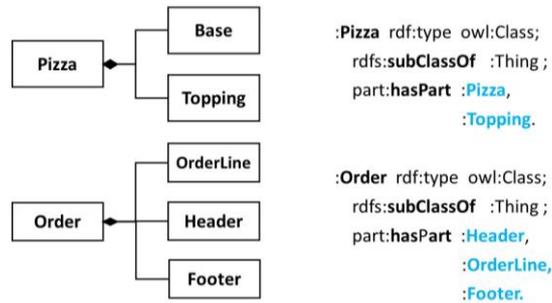

**Fig. 2**: Excerpt of structural PizzaShop Class Diagram and ontology fragment

Then in Fig. 3 an example of functional class diagram, together with its ontology fragment, is reported. Please note the coloured boxes that represent actors.

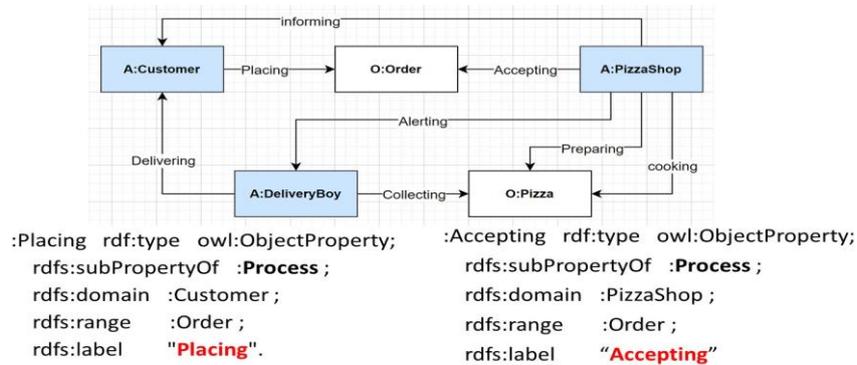

**Fig. 3**: Excerpt of functional PizzaShop Class Diagram and ontology

A formal representation offers various advantages, from the possibility of querying the knowledge artefact (e.g., to discover which actors perform what actions) to the possibility to apply a reasoner (we adopted Protégé) to prove the absence of (formal) inconsistencies (to this end, and to improve the models, constraints are added).

As anticipated, we presented the knowledge artefacts in a sequence, but in building them we proceed in a spiral way and at each step we look backward to guarantee the coherence of the different models. For instance, all the labels used in the CDs need to be already identified and reported in the Glossary. In the case that, when drawing a CD, new terms should emerge, we go back to the Glossary and the OPAAL Lexicon adding the new terms to them, in order to keep the different models aligned. Periodically, according to the Agile philosophy [12], the produced artefacts are released and shared with end-users and stakeholders for a validation. Then, comments and observations are used to improve the models and release a new version of the PBA Canvas knowledge base.

## 5   Related work

The area of business process analysis is very active, both at scientific and at industrial level, however knowledge-based BPA, that is the focus of this paper, presents only few results. Here we briefly review some of the key results in the area, with a focus on knowledge-based BPA and in particular on a BPA ontology [25].

In the quest for a formal method for BPA a few ontology-based solutions have been proposed. We may recall COBRA, a Core Ontology for Business pRocess Analysis [13] that is based on a Time Ontology. Another research line, with a wider scope, is represented by the adoption of ontologies and semantic web services for BP management, such as Semantic Business Process Management (SBPM) [14]. Such proposals appear to be more inclined towards the formal aspects than the ease of use for business experts.

A different research line, rooted in the business culture, starts from an international business standard, the Universal Business Language (UBL) [15]. In essence, UBL is an open library of information components, such as Address, Price, Quantity, and business templates of the most common documents, such as Invoice, Order, Receipt, plus a number of standard process models. An interesting proposal [16] is based on the association of a business ontology to UBL, proposing some formal models of the UBL components and templates, including the UBL process flows. The formal implementation of the UBL ontology has been achieved by using OWL (the W3C Web Ontology Language). Probably due to an excess of formalization, the proposal was not accepted by the business community.

Another interesting proposal is represented by BPMO [17], a Business Process Modeling Ontology that besides UBL considers also other business modeling standards, including ebXML[2]. BPMO has been mainly conceived for interoperability, i.e., to allow the exchange of information among cooperating enterprises, rather than to support business process analysis.

Finally, it is worth to return to the mentioned Business Model Canvas [4] that inspired the BPA Canvas layout. The former addresses a high level, enterprise space with respect to our proposal that is focussed on business processes. Furthermore, the former remains at an informal level and lacks of a systematic approach for the modeling practices and the produced documents. Along this line, another work to be mentioned is the Business Process Canvas [18] that has a similar scope to our since it is conceived to support BP analysis. In this proposal, the only similarity is the idea of adopting the canvas to analyse business processes. Then, we can see various differences: firstly, it does not intend to produce a formally grounded knowledge-base, in fact the gathered domain knowledge is represented by informal descriptions. Then, the organization of the canvas sections is very different from our BPA Canvas and there are not systematic guidelines and rules that suggest how to proceed. Finally, the BP Canvas has been mainly conceived for different purposes (e.g., interoperability, decision making), while modeling is a marginal objective, conversely, BPA Canvas has been primarily conceived for knowledge acquisition and modeling. Finally, BP Canvas still requires a field validation. On the contrary, BPA Canvas is currently going through two field tests, in the area of SMEs (a fashion atelier), and Public Administration (Italian Ministry of Economy and Finance). The first feedback is very encouraging.

In conclusion, in the literature there is a growing awareness of the importance of a knowledge-driven approach to BP analysis, but the existing proposals had a limited practical impact, failing in the objective of convincing business experts to adopt more rigorous and formal business process modeling methods. As anticipated, there are several causes: firstly, the clash of the business and the ontology cultures, with the pragmatism of the former and the formal approach of the latter. Then, the challenge in building large, encompassing, enterprise ontologies that turned out to be too complex, difficult to be achieved and maintained over time. We believe that starting with local solutions, e.g., a departmental or an application ontology, would have more chances of success. Also, the idea of pushing extensive competencies of ontology principles and the-ories in the business world appears not practical. There is a need for a 'soft' methodology that supports business experts with a progressive approach, from informal to formal, to knowledge modeling. And this is what characterise the BPA Canvas methodology.

---

[2] ebXML: Electronic Business using eXtensible Markup Language, an international standard aimed at representing business concepts [19].

## 6      Conclusions and discussion

In this paper we presented the BPA Canvas, a methodology for the acquisition, modeling and management of business process knowledge. The knowledge organization is guided by the canvas structured in eight sections that represents a sort of knowledge dashboard providing a synoptic view of the BPA models. As seen, with respect to previous proposals in the area of BPA, this methodology presents three key characteristics: (i) it is conceived to put the business experts at the center of the analysis process; (ii) it is characterised by a progression of model building, from informal to formal, that facilitates business experts in assuming a central role; (iii) it is rooted in formal knowledge management with a BPA Ontology. The methodology aims at modeling the structural knowledge, postponing the behavioural modeling, i.e., the diagramming of the process workflow, to the next design phase.

It is important to emphasise once more that out of eight knowledge artefacts the first seven can be easily built by business experts without the need of advanced technical competences. Only the final artefact, the BPA Ontology, requires specific knowledge engineering competencies. We believe that giving to business experts a central role has a number of advantages, first of all, as anticipated, it contributes to solve the mentioned Business/IT alignment problem. Then, the proposed knowledge management approach appears easy to be adopted also by SMEs that, traditionally, lack of competencies and resources required to carrying out innovation, supported by advanced methodologies [20]. On a more technical ground, the BPA Ontology, and the associated semantic services (e.g., semantic search, automatic reasoning, etc.), are fundamental to achieve a high quality business process analysis. Finally, with the emergence of the Low Code technology [21], coding activities are losing ground in favour of more high level business modeling. Then, the BPA Canvas, or any other rigorous BP design methodology, represents a virtuous pathway towards Low Code information systems development [22].

Our work will continue along two main lines. The first intends to evolve the BPA Canvas to bridge the analysis phase with the BP design phase. Then the next knowledge artefacts will be aimed to achieve a full-fledged business process flow, with the sequencing of activities and tasks and gateways with decision points. In particular, we are experimenting the adoption of the international standard OMG-BPMN (BP Modeling and Notation) [23].

The second line is represented by the development of a digital platform aimed at supporting business experts in building the BPA Canvas knowledge artefacts. The platform will offer various services, from the support to knowledge acquisition to the check of alignment and consistency of the different models, today such operations need to be achieved manually.

About knowledge acquisition, another interesting service is a chatbot capable to interviewing the business experts to build the first three artefacts (BP Signature, Statement, and User Story). Then, it is possible to proceed with the automatic extract of the terminology to populate the Glossary and the OPAAL Lexicon. A preliminary solution based on Wikidata is currently under development and it looks promising. The third service is the support in building the various Class Diagrams, starting from the Lexicon, and finally, a fourth service is aimed at building of the BPA Ontology starting from the CD (we are currently analyzing the rich literature on this topic, e.g. OntoUML [26]).

The work presented in this paper is the continuation of the work carried out in the context of the European Project BIVEE (Business Innovation in Virtual Enterprise Environment) where a first proposal of knowledge-based enterprise analysis has been proposed [24].